\documentstyle[prl,aps,preprint]{revtex}

\newcommand{\be}{\begin{equation}}
\newcommand{\ee}{\end{equation}}
\newcommand{\ba}{\begin{eqnarray}}
\newcommand{\ea}{\end{eqnarray}}
\newcommand{\bec}{\begin{center}}
\newcommand{\eec}{\end{center}}

\let\ssection=\section
\renewcommand{\section}{\setcounter{equation}{0}\ssection}

\begin{document}
\draft


\title{Multipole solutions in metric--affine gravity}

\author{Jos\'e Socorro$^{\$}$\thanks{E-mail: socorro@ifug4.ugto.mx},
Claus L\"ammerzahl$^\star$\thanks{E--mail:claus@spock.physik.uni-konstanz.de},
Alfredo Mac\'{\i}as$^\diamond$\thanks{E-mail: amac@xanum.uam.mx},
and Eckehard W. Mielke$^\diamond$\thanks{E-mail: ekke@xanum.uam.mx}\\ 
$^{\$}$ Instituto de F\'{\i}sica de la Universidad de Guanajuato,\\
Apartado Postal E-143, C.P. 37150, Le\'on, Guanajuato, MEXICO.\\
$\star$ Fakult\"at f\"ur Physik,
Universit\"at Konstanz,\\
Postfach 5560 M674, D--78434 Konstanz, GERMANY.\\ 
$^{\diamond}$ Departamento de F\'{\i}sica,\\
Universidad Aut\'onoma Metropolitana--Iztapalapa,\\
Apartado Postal 55-534, C.P. 09340, M\'exico, D.F., MEXICO.
}

\date{\today}

\maketitle

\begin{abstract}
Above Planck energies, the spacetime might become non--Riemannian, 
as it is known fron string theory and inflation. Then geometries arise in 
which nonmetricity and torsion appear as field strengths, side by side
with curvature. By gauging the affine group, a metric affine gauge theory 
emerges as dynamical framework. Here, by using the harmonic map ansatz, a new 
class of multipole like solutions in the metric affine gravity theory (MAG)  
is obtained. 

\end{abstract}

\pacs{PACS numbers:0450,0420J, 0350K}



\section{Introduction}

The study of the gravitational interaction coupled to Maxwell and  
dilaton fields has been the subject of recent investigations. In the low  
energy limit of string theory and as a result of dimensional reduction of  
the Kaluza--Klein Lagrangian, there arises the following four dimensional 
effective action \cite{garho}:
\begin{equation}
S = \int  \left[ -\frac{1}{2\kappa}R^{\alpha\beta} \wedge
\eta_{\alpha\beta} + 2 D\Phi \wedge ^*D\Phi  
- \frac{1}{2} e^{-2\alpha \Phi} F\wedge ^*F \right] 
\label{stemkk}\, ,
\end{equation}
where we denote Einstein's gravitational constant by 
$\kappa=\ell^2/(\hbar c)$, with $\ell$ the Planck length,, $\Phi$ is  
the 
scalar dilaton field, $F= dA$ is the Faraday electromagnetic field  
strength and $\alpha$ is the dilaton coupling constant that distinguishes the  
special subtheories contained in (\ref{stemkk}). For $\alpha=0$ we have the  
effective action of the Einstein--Maxwell--dilaton theory, $\alpha=1$  
represents the low energy string theory where only the $U(1)$--vector gauge 
field has not been dropped out, and, for $\alpha = \sqrt{3}$ the action  
(\ref{stemkk}) is an effective model obtained via the dimensional reduction 
of the corresponding five--dimensional Kaluza--Klein theory.
For all the corresponding field equation solutions are known  
\cite{mama}. 
In addition, a scheme has been developed in order to generate  
solutions of this model from solutions of the Laplace equation\cite{nk,mama}. 

In this paper we are going to map this class of solutions to  
solutions of metric--affine gravitational theories. 
The geometrical quantities in a metric--affine gravitational theory  
(MAG) are a pseudo--Riemannian metric, a coframe and a connection which may  
possess post--Riemannian structures, namely torsion and  
non--metricity (for a survey of these theories see \cite{PR}). 
For  restricted irreducible pieces of torsion and non--metricity there are  
similarities between the Einstein--Maxwell system and the MAG field  
equations. 
This observation enables us to find new solutions for MAG theories. 
In order to arrive at new solutions, we impose the condition of  
stationarity and spherical symmetry.  

A new solution with these symmetries with an additional  
electromagnetic field of a point charge has been presented in  
\cite{PLH97}. 
The results of the latter work was to confirm the general  
structure that the electromagnetic field is not directly influenced  
by the post--Riemannian structures torsion and non--metricity. 
In this spirit, we look for a wider class of solutions of the MAG field  
equations alone. 


\section{Harmonic map ansatz \label{HarmonicMapAnsatz}}

To begin with we consider the Papapetrou metric in the following 
parametrization:
\begin{equation}
ds^2 = - {1\over f^2}dt^{2} + f^2\left[ e^{2k}(d\rho^{2} 
+ d\zeta^{2}) + \rho^{2} d\varphi^{2} \right] 
\label{PPP}\, .
\end{equation}
In this Papapetrou parametrization, the Einstein--Maxwell field  
equations corresponding to the Lagrangian (\ref{stemkk}) reduce to  
the following set of equations:
\begin{eqnarray}
\Delta \ln f & = & e^{-2\alpha\Phi}{1\over\rho}f\  
A_{\varphi,z}A_{\varphi,\bar z} \label{feq} \, ,\\
2k_{,z} & = &  
4\rho(\Phi_{,z})^2-e^{-2\alpha\Phi}{f\over\rho}(A_{\varphi,z})^2+  
\rho\ (\ln f_{,z})^2 \label{keq}\, ,\\
2k_{,\bar z} & = & 4\rho(\Phi_{,\bar z})^2-e^{-2\alpha\Phi}{f\over\rho}
(A_{\varphi,{\bar z}})^2+  
\rho\ (\ln f_{,\bar z})^2 \label{keqb}
\end{eqnarray}
for the functions $f$ and $k$, respectively.

The {\em harmonic map ansatz} supposes that each quantity appearing  
in the metric, namely $f$ and $k$, depends on a set of functions  
$\lambda^i(z)$, ($i = 1, \ldots, p$ for some integer $p$): $f =  
f(\lambda^i)$ and $k = k(\lambda^i)$.  Here each function  
$\lambda^i(z)$ is assumed to fulfill the Laplace equation  
\cite{nk,mai}
\begin{equation}
\Delta\lambda^i=(\rho \lambda^i_{,z})_{,\bar z} 
+ (\rho \lambda^i_{,\bar z})_{,z} = 0 
\label{hmap}\, , 
\end{equation}
where 
\begin{equation} 
z = \rho + i\zeta
\label{eq:zet}\, .
\end{equation}
Thus the field equations derived from the Lagrangian (\ref{stemkk}) 
transform to equations in terms of the functions $\lambda^i$. 
In general these equations are easier to solve than the original  
ones. 

Another very important advantage of this method is based on the fact  
that it is
possible to generate for each solution of the Laplace equation  
(\ref{hmap}) an exact solution of the field equations derived from  
(\ref{stemkk}). 

A further fortunate feature of this method is that the harmonic map  
determines the gravitational and the electromagnetic potentials in  
such a way, that we can choose them to have electromagnetic  
monopoles, dipoles, quadrupoles, etc. \cite{mama}. It is important to note
that we have only two constants of integration comming from the solutions
to the Laplace equation. One of them is used to guarantee that the solution 
\cite{nk} is regular, and the other one remains free.

Let us now suppose that $\alpha = 0$ and that the components of the  
Papapetrou metric depend on one harmonic map $\lambda$ only. 
For the corresponding field equations
(\ref{feq}), (\ref{keq}) and (\ref{keqb}) we can present a solution: 
The field equation (\ref{keqb}) for the function $k$ is always  
integrable if 
$\lambda$ is a solution of the Laplace equation (for more details of 
this method see \cite{mat}).
The class of solutions we want to deal with here contains the  
electromagnetic field of a point charge $q$ and is given by
\begin{equation}
f = (1 - \lambda)^2,\quad  
k=0,\quad A_{\varphi,z} = q\rho \lambda_{,z},\quad  A_{\varphi,\bar z} = - q
\rho \lambda_{,\bar z}
\label{sol2}\, .
\end{equation}


\section{General quadratic MAG Lagrangian}

In a metric--affine spacetime, the curvature has {\em eleven} irreducible
pieces, see \cite{PR}, Table~4.  If we recall that the nonmetricity
has {\em four} and the torsion {\em three} irreducible pieces, then a  
general quadratic Lagrangian in MAG reads:
\begin{eqnarray} 
\label{QMA} V_{\rm MAG}&=&
\frac{1}{2\kappa}\,\left[-a_0\,R^{\alpha\beta}\wedge\eta_{\alpha\beta} 
-2\lambda\,\eta+  
T^\alpha\wedge{}^*\!\left(\sum_{I=1}^{3}a_{I}\,^{(I)}
T_\alpha\right)\right.\nonumber\\
&+&\left.  2\left(\sum_{I=2}^{4}c_{I}\,^{(I)}Q_{\alpha\beta}\right)
\wedge\vartheta^\alpha\wedge{}^*\!\, T^\beta + Q_{\alpha\beta}
\wedge{}^*\!\left(\sum_{I=1}^{4}b_{I}\,^{(I)}Q^{\alpha\beta}\right)\right]
\nonumber\\&- &\frac{1}{2}\,R^{\alpha\beta} \wedge{}^*\!
\left(\sum_{I=1}^{6}w_{I}\,^{(I)}W_{\alpha\beta} +
  \sum_{I=1}^{5}{z}_{I}\,^{(I)}Z_{\alpha\beta}\right)
\label{lobo}\,.  
\end{eqnarray} 
In the above, the signature of spacetime is $(-+++)$, $\eta:={}^*\!\, 1$ is 
the volume four--form and the constants $a_0,\cdots a_3$, 
$b_1,\cdots b_4$, $c_2, c_3,c_4$,  
$w_1,\cdots w_6$, $z_1,\cdots z_5$ are dimensionless. In the curvature square  
term we have introduced the irreducible pieces of the antisymmetric part
$W_{\alpha\beta}:= R_{[\alpha\beta]}$ and the symmetric part
$Z_{\alpha\beta}:= R_{(\alpha\beta)}$ of the curvature two--form.
Again, in $Z_{\alpha\beta}$, we meet a purely post--Riemannian part.
The segmental curvature $^{(4)}Z_{\alpha\beta}:=
R_\gamma{}^\gamma\,g_{\alpha\beta}/4= g_{\alpha \beta} dQ$ has  
formally a 
similar structure as the electromagnetic field strength $F=dA$, but is 
physically quite different since it is related to Weyl rescalings.  
 
Let us recall the three general field equations of MAG,
see \cite{PR} Eqs.(5.5.3)--(5.5.5). Because of its redundancy, we can omit
the zeroth field equation with its gauge momentum $M^{\alpha\beta}$.  
The first and the second field equations read 
\begin{eqnarray} 
DH_{\alpha}- E_{\alpha}&=&\Sigma_{\alpha}\,,\label{first}\\ 
DH^{\alpha}{}_{\beta}- E^{\alpha}{}_{\beta}&=&\Delta^{\alpha}{}_{\beta}\,,
\label{second}
\end{eqnarray} 
where the three--forms $E_{\alpha}$ and $E^{\alpha}{}_{\beta}$
describe the canonical energy--mo\-men\-tum and hypermomentum  
currents of the gauge fields themselves, whereas
$\Sigma_{\alpha}$ and $\Delta^{\alpha}{}_{\beta}$ are the
canonical energy--momentum and hypermomentum current three--forms
associated with matter. Here we will consider only the {\em vacuum case} where
$\Sigma_{\alpha}=\Delta^{\alpha}{}_{\beta}=0$. The left hand sides of
(\ref{first})--(\ref{second}) involve the gravitational gauge field
momenta two-forms $H_{\alpha}$ and $H^{\alpha}{}_{\beta}$
(gravitational ``excitations"). We find them, together with  
$M^{\alpha\beta}$, by partial differentiation of the Lagrangian (\ref{QMA}):
\begin{eqnarray}
M^{\alpha\beta}&:=&-2{\partial V_{\rm MAG}\over \partial  
Q_{\alpha\beta}}=
-{2\over\kappa}\Bigg[{}^*\! \left(\sum_{I=1}^{4}b_{I}{}^{(I)}
Q^{\alpha\beta}\right)\nonumber\\
&& + c_{2}\,\vartheta^{(\alpha}\wedge{}^*\! ^{(1)}T^{\beta)} +
c_{3}\,\vartheta^{(\alpha}\wedge{}^*\! ^{(2)}T^{\beta)} +
{1\over 4}(c_{3}-c_{4})\,g^{\alpha\beta}{}^*\!\,   
T\Bigg]\,,\label{M1}\\
  H_{\alpha}&:=&-{\partial V_{\rm MAG}\over \partial T^{\alpha}} = -
  {1\over\kappa}\,
  {}^*\!\left[\left(\sum_{I=1}^{3}a_{I}{}^{(I)}T_{\alpha}\right) +
    \left(\sum_{I=2}^{4}c_{I}{}^{(I)}
Q_{\alpha\beta}\wedge\vartheta^{\beta}\right)\right],\label{Ha1}\\
  H^{\alpha}{}_{\beta}&:=& - {\partial V_{\rm MAG}\over \partial
    R_{\alpha}{}^{\beta}}= {a_0\over  
2\kappa}\,\eta^{\alpha}{}_{\beta} +
  {\cal W}^{\alpha}{}_{\beta} + {\cal Z}^{\alpha}{}_{\beta},\label{Hab1}
\end{eqnarray}
where we introduced the abbreviations
\begin{equation}
  {\cal W}_{\alpha\beta}:= {}^*\!
  \left(\sum_{I=1}^{6}w_{I}{}^{(I)}W_{\alpha\beta} \right),\quad\quad
  {\cal Z}_{\alpha\beta}:= {}^*\!
  \left(\sum_{I=1}^{5}z_{I}{}^{(I)}Z_{\alpha\beta} \right).
\end{equation}


\section{Harmonic field configuration in MAG}

We work in the spherical polar coordinates $(t,r,\theta,\phi)$, which  
lead  
to the {\em isotropic} coframe 
\begin{equation}
  \vartheta ^{\hat{0}} =\,{1\over f}\, d\,t \,,\quad
\vartheta ^{\hat{1}} =\, f\, d\, r\, , \quad
\vartheta ^{\hat{2}} =\, f\, r\, d\,\theta\,,\quad
\vartheta ^{\hat{3}} =\, f\, r\, \sin\theta \, d\,\phi
  \,,\label{frame2}
\end{equation}
with one unknown function $f=f(r,\theta)$. Since the coframe is assumed to  
be {\em orthonormal} with the local Minkowski metric
$o_{\alpha\beta}:=\hbox{diag}(-1,1,1,1) =o^{\alpha\beta}$, we have  
the spherically symmetric metric in {\em isotropic} form 
\ba 
ds^2=o_{\alpha\beta}\,\vartheta^\alpha\otimes\vartheta^\beta 
&=& -\frac{1}{f^2}\,dt^2+ {f^2} \left[dr^2
+r^2\left(d\theta^2+\sin^2\theta \,d\phi^2\right)\right]
\label{gps}\, .
\ea

As for the torsion and nonmetricity configurations, we concentrate on
the simplest non--trivial case with shear. According to its  
irreducible decomposition (see the Appendix B of \cite{PR}), the nonmetricity
contains two covector pieces, namely $^{(4)}Q_{\alpha\beta}=
Q\,g_{\alpha\beta}$, the dilation piece, and
\begin{equation}
  ^{(3)}Q_{\alpha\beta}={4\over
    9}\left(\vartheta_{(\alpha}e_{\beta)}\rfloor \Lambda - {1\over
      4}g_{\alpha\beta}\Lambda\right)\,,\qquad \hbox{with}\qquad
  \Lambda:= \vartheta^{\alpha}e^{\beta}\rfloor\!
  {\nearrow\!\!\!\!\!\!\!Q}_{\alpha\beta}\label{3q}\,,
\end{equation}
a proper shear piece. Accordingly, our ansatz for the nonmetricity  
reads
\begin{equation}
  Q_{\alpha\beta}=\, ^{(3)}Q_{\alpha\beta} +\,
  ^{(4)}Q_{\alpha\beta}\,.\label{QQ}
\end{equation}
The torsion, in addition to its tensor piece,
encompasses a covector and an axial covector piece. Let us choose only
the covector piece as non-vanishing:
\begin{equation}
T^{\alpha}={}^{(2)}T^{\alpha}={1\over 3}\,\vartheta^{\alpha}\wedge  
T\,,
\qquad \hbox{with}\qquad T:=e_{\alpha}\rfloor T^{\alpha}\,.\label{TT}
\end{equation}
Thus we are left with the three non--trivial one--forms $Q$,  
$\Lambda$, and $T$. 

In the spherically symmetric case, they should not distinguish
a direction in space. The following ansatz turns to be compatible  
with that 
condition,\footnote{In the projective invariant Einstein Lagrangian  
$V(R)$, 
there arise the projective invariant $T+(3/2)Q= (3/2)d\ln V^\prime$  
cf. Mac\'{\i}as et al. \cite{mamimo}, which seems to surface in a related  
way in the coefficient of (\ref{b4}).}
\begin{equation}
Q=k_0  
\frac{u(r,\theta)}{r}\,\vartheta^{\hat{0}}=\frac{k_0}{k_1}\Lambda
=\frac{k_0}{k_2}T 
\label{genEug}\, .
\end{equation}
Here we introduced the second function $u(r, \theta)$ which has to be  
determined by the field equations of MAG. 
What is the geometrical content of the above ansatz, or, in other  
words, what are the consequences for the {\em curvature}? 
If we take the trace of the zeroth Bianchi identity 
\begin{equation}
DQ_{\alpha\beta} = 2 Z_{\alpha\beta} 
\label{0Bianchi}\, ,
\end{equation} 
it merely consists of one irreducible piece $2dQ = Z_\gamma{}^\gamma  
= \,^{(4)}Z_\gamma{}^\gamma$. 
Consequently $Q$ serves as a {\em potential} for
$^{(4)} Z_\gamma{}^\gamma$ in the same way as $A$ for $F=dA$. 
In addition, the third part of (\ref{0Bianchi}) reads
$^{(3)}(DQ_{\alpha\beta}) = 2\,^{(3)}Z_{\alpha\beta}$, where
\begin{equation}
  ^{(3)}Z_{\alpha\beta}={2\over 3}\left(\vartheta_{(\alpha}\wedge
    e_{\beta)}\rfloor\Delta - {1\over
      2}g_{\alpha\beta}\Delta\right)\,,
  \quad\hbox{with}\quad\Delta:={1\over 2}\vartheta^{\alpha}\wedge
e^{\beta}\rfloor\!{\nearrow\!\!\!\!\!\!\!Z}_{\alpha\beta}\,.\label{3z} 
\end{equation}
The similarity in structure of (\ref{3q}) and (\ref{3z}) is apparent.
Indeed, {\em provided} the torsion carries only a covector piece, see
(\ref{TT}), we find 
\begin{equation}
\Delta={1\over 6}\,d\Lambda\,,\label{ddl}
\end{equation}
i.e.\ $^{(3)}Q_{\alpha\beta}$ acts as a potential for
$^{(3)}Z_{\alpha\beta}$.


\section{Exact non--singular MAG solutions}

When we substitute the local metric $o_{\alpha\beta}$, the coframe
(\ref{frame2}), and the ansatz (\ref{genEug}) of the nonmetricity and  
torsion into the
field equations (\ref{first}), (\ref{second}) of the Lagrangian
(\ref{lobo}), we find that the function 
$f(r,\theta)$ has to satisfy the 2D Laplace equation
\begin{equation}
\frac{\partial}{\partial r}\left( r^2 \frac{\partial f}{\partial r}\right)+
\frac{1}{\sin \theta} \frac{\partial}{\partial \theta} \left(  
\sin\theta \frac{\partial f}{\partial \theta} \right) = 0\,  .
\end{equation}
Moreover, the function $u(r,\theta)$ in the ansatz (\ref{genEug}) has  
to be a solution of the following equation
\be
 r^2 \frac{\partial^2 u}{\partial r^2} +
\frac{1}{\sin \theta} \frac{\partial}{\partial \theta} \left(  
\sin\theta 
\frac{\partial u}{\partial \theta} \right) = 0
\label{legendre}\,  .
\ee
The general solutions are given by
\begin{eqnarray}
f(r,\theta) & = & \sum_{n = 0}^\infty \left(\widetilde Q_n^{(1)} r^n  
+ \widetilde Q_n^{(2)} {1\over{r^{n + 1}}} \right) P_n(\cos\theta)  \\ 
u(r,\theta) & = & \sum_{n = 0}^\infty \left(\widehat Q_n^{(1)} r^{n + 1} 
+ \widehat Q_n^{(2)} {1\over{r^n}} \right) P_n(\cos\theta)
\end{eqnarray}
Here $\widetilde Q_n^{(1)}$, $\widetilde Q_n^{(2)}$, $\widehat  
Q_n^{(1)}$, and $\widehat Q_n^{(2)}$ are arbitrary 
{\em integration constants}, and $P_n$ are the Legendre polynomials. 
Notice that the integration constants for $u(r,\theta)$ are embbeded in the 
constants appearing in the ansatz (\ref{genEug}).

The coefficients $k_{0}, k_{1}, k_{2}$ in
the ansatz (\ref{genEug}) are determined by the 
dimensionless coupling constants of the Lagrangian:
\ba
k_0 &=& \left({a_2\over 2}-a_0\right)(8b_3 + a_0) - 3(c_3 + a_0 )^2\,,
\label{k0}\\
k_1 &=& -9\left[ a_0\left({a_2\over 2} - a_0\right) + 
(c_3 + a_0 )(c_4 + a_0 )\right]\,,
\label{k1}\\
k_2 &=& {3\over 2} \left[ 3a_0 (c_3 + a_0 ) + (
8b_3 + a_0)(c_4 + a_0 )\right]\,.\label{k2}
\ea
A rather weak condition, which must be imposed on these coefficients, 
prescribes the value 
\begin{equation}
  b_4=\frac{a_0k+2c_4k_2}{8k_0}\,,\qquad\hbox{with}\qquad k:=
  3k_0-k_1+2k_2
\label{b4}\, .
\end{equation}
for the coupling constant $b_4$, and the following relation for $z_4$
\be
{\tilde Q}_n^2=\kappa z_4 \frac{(k_0 A)^2}{2a_0}
\label{z4}\, .
\ee
Due to the fact that the same function $u$ determines the triplet
of nonmetricity and torsion one--forms,
all constants of the series are equal, certainly a drawback of the ansatz 
(\ref{genEug}).

If we collect our results, then the nonmetricity and the torsion read
as follows:
\begin{eqnarray}
Q^{\alpha\beta} & = & \sum_{n = 0}^\infty \left(\widetilde Q_n^{(1)}  
r^n + \widetilde Q_n^{(2)} {1\over{r^{n + 1}}} \right)  
P_n(\cos\theta)\,\left[k_0 A\,o^{\alpha\beta}
+\frac{4}{9}\,k_1 A\,\left(\vartheta^{(\alpha}e^{\beta)}\rfloor-\frac{1}{4}\,
o^{\alpha\beta}\right)\right]\vartheta^{\hat{0}}  
\label{nichtmetrizitaet} \, ,\\ 
T^\alpha & = & \frac{k_2 A}{3}\,\sum_{n = 0}^\infty \left(\widehat  
Q_n^{(1)} r^n + \widehat Q_n^{(2)} {1\over{r^{n + 1}}} \right)  
P_n(\cos\theta)
\vartheta^\alpha\wedge\vartheta^{\hat{0}}
\,.\label{torsion1}
\end{eqnarray}
Since nonmetricity and torsion are geometric objects like curvature,  
infinities in their components at spatial infinity are non--physical.  
For example, torsion can be measured by spin--precession  
\cite{Laem}. An infinite torsion will then give rise to a spin  
precession with an infinite angular velocity at spatial infinity. 

In order to avoid such unphysical singularities, 
we have to demand the  
vanishing of the corresponding coefficients $\widetilde Q_n^{(1)} =  
0$ and $\widehat Q_n^{(1)} = 0$. 

Therefore, our final solution is
\begin{eqnarray}
Q^{\alpha\beta} & = & \sum_{n = 0}^\infty \widetilde Q_n^{(2)}  
{1\over{r^{n + 1}}}  P_n(\cos\theta)\,\left[k_0 A\,o^{\alpha\beta}
+\frac{4}{9}\,k_1 A\,\left(\vartheta^{(\alpha}e^{\beta)}\rfloor-\frac{1}{4}\,
o^{\alpha\beta}\right)\right]\vartheta^{\hat{0}}  
\label{nichtmetri} \\ 
T^\alpha & = & \frac{k_2 A}{3}\,\sum_{n = 0}^\infty \widehat  
Q_n^{(2)} {1\over{r^{n + 1}}} P_n(\cos\theta)  
\vartheta^\alpha\wedge\vartheta^{\hat{0}}
\,.\label{torsion2}
\end{eqnarray}
with free coefficients $\widetilde Q_n^{(2)}$ and $\widehat  
Q_n^{(2)}$ respresenting its multipolar structure.

This solution was checked with Reduce \cite{REDUCE} with its Excalc package 
\cite{EXCALC} for treating exterior differential forms\cite{Stauffer} and 
the Reduce--based GRG computer algebra system \cite{GRG}.


\section{Outlook}

On should be cautious that the physical motivation to go beyond 
classical Einstein gravity in the MAG model is not completely clear and well 
founded\cite{nehe}. 
Although in view of the problems of other theories, like supergravity and 
even string {\em field theory}\cite{ne97} in this respect,
it appears unfair to ask the question of renormalizability for MAG,
one would like to know at which energy scale such a framework 
can be regarded as an {\em effective} gravitational model. In Ref. \cite{PR}, 
the motivation for MAG came mainly from particle physics 
and the manifield description of an infinite tower of fermions. 
One may regards such gauge theories of gravity with Weyl invariance 
as some small but perspective step  towards quantum gravity.

Moreover, the nonmetricity one--form $=Q_{\alpha\beta}=
Q_{\alpha\beta i}dxŒ$ with its three spacetime components 
is known to introduce in general a spin 3 mode. So one needs to derive
restrictions on the parameters of the MAG Lagrangian to avoid {\em causality 
problems} which otherwise could occur in quantum field  theory \cite{Kaku} 
arising for such high spin fields.
In our case we restricted ourselves to an ansatz (\ref{QQ}) involving only 
two covector pieces in the nonmetricity with spin 1. Moreover, it is well 
known that Einstein's general relativity theory is satisfactorily supported 
by experimental tests on the macroscopic level. Thus, whereas the 
gravitational gauge models provide an alternative description of gravitational
physics, it is natural to require their correspondence with general relativity
at large distances. Unfortunately, direct generalization of the standard 
Hilbert--Einstein Lagrangian yields an unphysical MAG model which is 
projectively invariant and, accordingly, imposes unphysical constraints 
on the matter sources. 

Another essential difficulty in the development of a dynamical scheme of 
MAG was, until recently, the lack of self--consistent models which
describe physical (quantum, semiclassical, or classical) sources of MAG 
possessing mass or energy-momentum and hypermomentum.
It was proposed\cite{TW} to take as the gravitational Lagrangian the sum 
of the (generalized) Hilbert--Einstein term and the square of the segmental
curvature (thus reviving the old proposals of \cite{Palatini}). Further 
extensions of this model, which include the quadratic invariants of torsion 
and nonmetricity, were investigated in the vacuum case in Refs. 
\cite{Tres14,Tres15,heh96,he96,De96}.

In a recent preprint\cite{Ob97} it has been 
demonstrated that the  MAG model considered here can reduced  
to an {\em effective} Einstein--Proca system. This is admittedly more
elegant than the proof  of Dereli et al., Ref.\cite{De96}. 
On the other hand, however, this could also imply that 
this special MAG model has problems with {\em redundant variables}.

In the case of restricted Poincar\'e gauge models (without nonmetricity),  
a similar reduction (there induced there via a double duality ansatz)
was based on the {\em teleparallelism equivalence}, see Baekler et 
al. Ref.  \cite{bm86}. However, it was shown by Lenzen \cite{{le84}}, and 
later confirmed in Ref. \cite{bm88} that then necessarily {\em free} functions
occur in exact torsion solutions. (The tentavive gauge fixing approach 
suggested there as a way out met considerable criticism.) 
Thus for the so--called ``viable" set there exist 
{\em infinite} many exact vacuum solutions 
which may indicate a physically problematic {\em degeneracy} of those 
models. 
Recent reports to rescue the initial value problem in PG  theory
by Hecht et al. \cite{he96} and the Refs. therein, seem not to be very 
conclusive, since the  problem of  free functions seems there to be ``swept 
under the carpet".

The related situation for MAG is not yet resolved, since again 
a teleparallelism type relation, see (5.9.16) of Ref. \cite{PR},
seems to be crucial for the equivalence proof of MAG with the Einstein--Proca 
Lagrangian. 


\section{\bf Acknowledgments}
We would like to thank Friedrich W. Hehl for valuable hints.
This research was supported by  CONACyT, grants No. 3544--E9311,
No. 3898P--E9608, and by the joint German--Mexican project  
KFA--Conacyt E130--2924 and DLR--Conacyt 6.B0a.6A.
Moreover, E.W.M. acknowledges the support by the 
short--term fellowship 961 616 015 6 of the German Academic Exchange  
Service (DAAD), Bonn, and C.L. thanks the Deutsche  
Forschungsgemeinschaft for financial support.


\begin{references}                    


\bibitem{garho} D. Garfinkle, G.T. Horowitz, and A. Strominger, 
{\em Phys. Rev.} {\bf D43} (1991) 3140.

\bibitem{mama} A. Mac\'{\i}as and T. Matos, {\em Class. Quantum Grav.}
{\bf 13} (1996) 345.

\bibitem{nk} G. Neugebauer and D. Kramer, {\em Ann. Phys.}, (Leipzig)  
{\bf 24} (1970) 62.

\bibitem{PR} F.W. Hehl, J.D. McCrea, E.W.  Mielke, and Y. Ne'eman,
{\em Phys. Rep.} {\bf 258} (1995) 1.

\bibitem{PLH97} R.A. Puntigam, C. L\"ammerzahl, F.W. Hehl,  
{\em Class. Quant. Grav.} {\bf 14} 1347.

\bibitem{mai} D. Maison, {\em Phys. Rev. Lett.} {\bf 41} (1978) 521.

\bibitem{mat} T. Matos, {\em J. Math. Phys.} {\bf 35} (1994) 1302.

\bibitem{mamimo} A. Mac\'{\i}as, E. W. Mielke, H.A. Morales--T\'ecotl, and
R. Tresguerres, {\em J. Math. Phys.} {\bf 36} (1995) 5868.

\bibitem{Laem} C. L\"ammerzahl, {\em Phys. Lett.} {\bf A228} (1997) 223.

\bibitem{REDUCE} A.C. Hearn, {\em REDUCE User's Manual. Version 3.6}.
  Rand publication CP78 (Rev.\ 7/95) (RAND, Santa Monica, CA
  90407-2138, USA, 1995).

\bibitem{EXCALC} E. Schr\"ufer, F.W. Hehl, and J.D. McCrea, {\em Gen. Relat. 
Grav.} {\bf 19} (1987) 197.

\bibitem{Stauffer} D. Stauffer, F.W. Hehl, N. Ito, V. Winkelmann, and 
J.G. Zabolitzky: {\em Computer Simulation and Computer Algebra --
Lectures for Beginners.} 3rd ed.\  (Springer, Berlin, 1993).

\bibitem{GRG} V.V. Zhytnikov, {\em GRG. Computer Algebra System for
Differential Geometry, Gravity and Field Theory. Version 3.1} (Moscow, 1991) 
108 pages.

\bibitem{nehe} Y. Ne'eman and F.W. Hehl, {\em Class. Quantum Grav.} Supl. 
{\bf 14} (1997) A251. 

\bibitem{ne97} Y. Ne'eman: ``Status of quantum gravity", in:
{\em Recent Developments in Gravitation and Mathematical Physics},
Proceedings of the $2^{nd}$ Mexican School on Gravitation and Mathematical
Physics, Tlaxcala, Tlax. 1996. A. Garc\'{\i}a et al. eds. (Science Network
Publishing 1997). 

\bibitem{Kaku}M. Kaku: {\em Quantum Field Theory} (Oxford University Press 
1993). p.  750.

\bibitem{TW} R.W.  Tucker and C.  Wang,  {\em Class. Quantum Grav.} 
{\bf 12} (1995) 2587.

\bibitem{Palatini} F.W. Hehl, E.A. Lord, and L.L. Smalley, 
{\em Gen.  Relat. Grav.} {\bf 13} (1981) 1037.

\bibitem{Tres14} R. Tresguerres, {\em Z. Phys.} {\bf C65} (1995) 347.

\bibitem{Tres15} R. Tresguerres, {\em Phys. Lett.} {\bf A200} (1995)  
405.

\bibitem{heh96}Yu.N. Obukhov, E.J. Vlachynsky, W. Esser, R.  
Tresguerres, and F.W. Hehl, {\em Phys. Lett.} {\bf A220} (1996) 1.

\bibitem{he96} E.J. Vlachynsky, R. Tresguerres, Yu.N. Obukhov, and  
F.W. Hehl, {\em Class. Quantum Grav.} {\bf 13} (1996) 3253.

\bibitem{Ob97}Yu.N.\ Obukhov, E.J.\ Vlachynsky, W.\ Esser, and F.W.\ Hehl: 
``Irreducible decompositions in metric-affine gravity models", preprint 
University of Cologne 1997.

\bibitem{De96}
T. Dereli, M. \"Onder, J. Schray, R.W. Tucker, and C. Wang, 
{\em Class. Quantum Grav.} {\bf 13} (1996) L103.


\bibitem{bm86} P. Baekler and E.W. Mielke, {\em Phys. Letters} {\bf 113A} 
(1986) 471.

\bibitem{le84} J. Lenzen, {\em Nuovo Cim.} {\bf B82} (1984) 85.

\bibitem{bm88} P. Baekler and E.W. Mielke, {\em Fortschr. Phys.} {\bf 36} 
(1988) 549.

\bibitem{he96} R.D. Hecht, J.M. Nester, and V.V. Zhytnikov, {\em Phys. Lett.}
{\bf A222} (1996) 37.

\end{references}
\end{document}